# Chemically homogeneous boron carbide $^{10}$B/$^{11}$B isotope modulated neutron interference mirrors

Jens Birch, Sjoerd Stendahl, Samira Dorri, Anton Zubayer, Naureen Ghafoor, Fredrik Eriksson*

*Department of Physics, Chemistry, and Biology, IFM, Linköping University, SE-581 83 Linköping, Sweden*

*Corresponding author: fredrik.eriksson@liu.se

**Abstract**

We introduce a novel type of neutron interference mirrors based on a chemically homogeneous $B_xC$ (x>4) matrix with internal high precision $^{10}$B/$^{11}$B isotope modulation. Simulations predict that these mirrors exhibit very high neutron reflectivities for a small number of bilayer periods. This is experimentally confirmed by neutron reflectivity measurements of mirrors synthesized by ion-assisted magnetron sputter deposition. For example, a 120 nm thick multilayer consisting of just 20 bilayer periods of $^{10}$B$_{5.7}$C/$^{11}$B$_{5.7}$C, with a periodicity of 61.5 Å, exhibits a neutron reflectivity of 13% at an incidence angle of 4.7° for neutrons with a wavelength of 4.825 Å. This is attributed to a high scattering length density contrast between the layers with an interface width <5Å. Structural analyses by X-ray diffraction, X-ray reflectivity, and transmission electron microscopy demonstrate that the $^{10}$B$_{5.7}$C/$^{11}$B$_{5.7}$C multilayer mirrors are composed of amorphous, chemically homogenous $B_{5.7}$C, without any internal chemical modulation. The data show that $^{10}$B$_x$C/$^{11}$B$_x$C multilayer mirrors have the potential for higher neutron reflectivities at higher q-values using fewer and thinner layers, compared to today's state-of-the-art chemically modulated neutron multilayer mirrors.

**Introduction**

The design of highly reflective (non-polarizing) multilayer neutron mirrors has so far generally followed three design criteria regarding the choice of layer materials: 1) High difference in scattering length densities (SLDs) to maximise reflectance in each interface. 2) Minimal absolute values of the imaginary parts of the SLDs to eliminate loss of neutrons due to absorption. 3) The materials should form abrupt and flat interfaces to maximise local SLD contrast, minimise loss of coherence between adjacent interfaces, and minimise loss of neutrons due to diffuse scattering. An example is Ni/Ti which is the materials system of choice for most supermirrors, satisfying criteria 1 and 2 reasonably well but failing on criterion 3, where reflectivity is hampered by interdiffusion and roughness evolution upon NiTi intermetallic interphase formation leading to interface widths no smaller than 0.7 nm.[1,2]

We introduce here the idea to eliminate any chemical driving force for deteriorating interface abruptness by employing an isotope modulation, rather than a chemical modulation, to obtain a high SLD contrast in multilayer mirrors. Chemically homogenous, isotope-modulated multilayered materials have been used for utilizing neutron scattering to study self-diffusion[3] in chemically homogeneous polycrystalline, as well as amorphous[4] multilayers, as well as to reduce phonon scattering in wide band-gap single crystal $^{12}$C/$^{13}$C diamond superlattices.[5] However, there exist no reports on $^{10}$B/$^{11}$B isotope-modulated multilayers, and we are not aware of any previous work utilizing isotope modulation in this way for the application of neutron mirrors.

Natural boron is composed of two isotopes $^{10}$B and $^{11}$B at a ratio 1:4 and thanks to their large relative mass difference isotope-enriched boron carbides, $^{10}$B$_4$C and $^{11}$B$_4$C, are relatively affordable materials. B$_4$C is a refractory compound with high stability that can be sputter deposited as dense smooth amorphous layers over large areas[6], exhibiting low film stress and high neutron radiation hardness.[7] Moreover, bulk boron carbide is known to exhibit a wide compositional range of stable $B_xC$ with B/C relative content x in the range $4 \leq x \leq 12.5$[8,9] implying that a higher SLD contrasts should be achievable in $^{10}$B$_x$C/$^{11}$B$_x$C multilayers if x>4. Therefore, $^{10}$B$_x$C/$^{11}$B$_x$C multilayers have the potential to be a suitable material for application in the field of neutron interference mirrors. As can be seen in Fig. 1, when taking into account both the real and imaginary parts of their SLDs, the difference in SLD for $^{10}$B and $^{11}$B is larger than that of Ni and Ti. When the boron isotopes are stabilized as boron carbides, $^{10}$B$_x$C and $^{11}$B$_x$C, the SLD-contrast will decrease as the relative B/C content x decreases from x=12.5 to x=4, where $^{10}$B$_4$C and $^{11}$B$_4$C provide an SLD contrast similar to that of Ni and Ti. This implies that a high neutron reflectance is possible from an amorphous $B_xC$ layer with internal $^{10}$B/$^{11}$B isotope modulation, thus fulfilling design criterion 1.

It should be noted that $^{10}$B$_4$C has a non-zero imaginary SLD which is manifested as a high absorption of neutrons. This means that design criterion 2 is not fully met. However, this will mainly be a limitation for devices, such as supermirrors, which rely on a large number of interfaces and operating at very grazing incidence angles with intrinsically long beam paths through absorbing $^{10}$B-containing layers. Thus, $^{10}$B/$^{11}$B isotope modulated multilayers will be most suited for non-grazing incidence devices such as band-pass mirrors and multilayer monochromators.

This materials system will fulfil design criterion 3, since the absence of any internal chemical gradient eliminates almost all detrimental effects for spoiling the SLD contrast between adjacent layers. Not only chemically driven effects, such as interdiffusion and interphase interlayer formation, will be absent, but also difficult-to-control kinetic effects such as preferential re-sputtering and sputter yield amplification[10,11,12,13,14] at the interfaces during sputter deposition of the multilayer mirror will be eliminated. Furthermore, the low atomic masses of B and C eliminate the issue of intermixing at the interfaces by back-reflected high energy neutrals.[13,15] Finally, realizing the $^{10}$B/$^{11}$B modulation in an amorphous boron carbide matrix will also eliminate interface roughness and diffuse neutron scattering due to nanocrystallites.

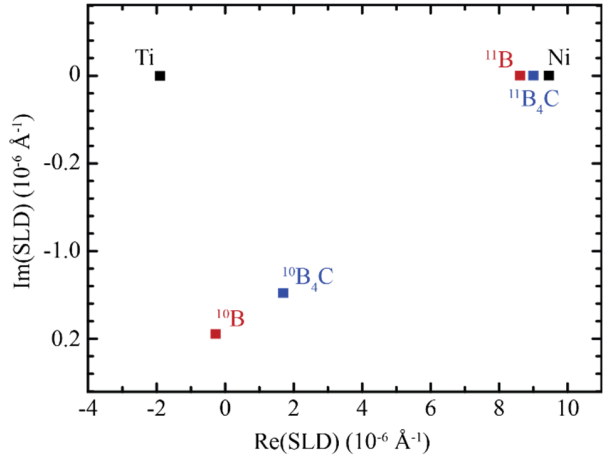

**Figure 1.** Real and imaginary parts of the scattering length densities (SLDs) of Ni, Ti, $^{10}$B, $^{11}$B, $^{10}$B$_4$C, and $^{11}$B$_4$C. See Experimental Details regarding determining SLDs.

**Experimental Details**

$^{10}$B/$^{11}$B isotope-modulated boron carbide mirrors were deposited using a high vacuum magnetron sputter deposition chamber at Linköping University, specially designed for ion-assisted interface engineering during deposition of neutron and X-ray multilayer mirrors.[16,17] The mirrors were grown on 10×10×0.5 mm$^3$ Si (001) substrates which were ultrasonically cleaned for 5 minutes in each of trichloroethylene, acetone, and isopropanol alcohol whereafter they were blown dry in N$_2$-gas just prior to being load-locked into the deposition chamber. The substrates were rotated around the sample normal at a constant rate of 17 rpm during the growth process. The samples were deposited at ambient temperatures with no substrate heating. High purity argon gas (>99.9997%) was used as sputtering gas at a working pressure of 3 mTorr (0.4 Pa), as measured by a capacitance manometer. The background base pressure of the chamber just prior to sample growth was less than to $5.0 \cdot 10^{-7}$ Torr (0.67 µPa) as achieved by a turbo molecular pump. The 75 mm-diameter $^{10}$B$_4$C and $^{11}$B$_4$C sputtering targets (purity >98,7%) were operated at constant power of 100 W, resulting in deposition rates of 0.192 Å/s and 0.198 Å/s for $^{10}$B$_4$C and $^{11}$B$_4$C, respectively, as measured by X-ray reflectivity. $^{10}$B$_4$C/$^{11}$B$_4$C multilayers were grown to a nominal thickness of 128 nm with $^{10}$B$_4$C+$^{11}$B$_4$C bilayer periods $\Lambda$ ranging from $\Lambda$=16 Å to $\Lambda$=128 Å and with $^{10}$B$_4$C layer thickness-to-period ratios $\Gamma$ ranging from $\Gamma$=0.125 to $\Gamma$=0.875.

The deposition of each individual layer of $^{10}$B$_4$C and $^{11}$B$_4$C was performed using a two-stage low energy ion assistance scheme in which the first 0.3 nm of each layer was grown without any ion assistance in order to protect the interfaces from ion-induced intermixing during their formation. In order to mitigate roughening due to kinetically limited growth, the remaining part of each layer was grown with a mild Ar ion-assistance by applying an attractive substrate voltage of -30 V. A solenoid was used to guide electrons towards the substrate table and create a denser plasma near the substrate surface to significantly increase the ion-to-adatom arrival rate ratio during the second, low energy ion assisted stage of growth of each layer. This allowed adatoms to migrate at least once after reaching their landing sites, while maintaining low ion energy to prevent bulk diffusion.[18] This interface engineering method for eliminating intermixing and reducing roughness has been successful for many multilayer mirror systems such as Ni/Ti,[17] Vi/V,[18] Cr/Sc,[19] and Ti/Cr.[20]

The elemental and isotopic compositions of the films were determined using time-of-flight elastic recoil detection analysis (ToF-ERDA) at the Tandem Laboratory at Uppsala University. A primary beam of $^{127}$I$^{8+}$ was used with an energy of 36 MeV at an incident angle of 67.5° relative to the surface normal, the energy detector was placed at a recoil scattering angle of 45°. A detailed description of the experimental set-up is given elsewhere.[21,22] The measured data has been analysed using the Potku software in order to determine the atomic concentrations.[23]

Neutron reflectometry (NR) measurements were performed at the neutron reflectometer MORPHEUS at the Swiss Spallation Neutron Source SINQ. Using a monochromatic wavelength λ=4.85 Å the reflectivity was measured as a function of scattering angle in the range 0-6º 2θ using a step size of 0.01º per step in 2θ with 20 seconds per step, resulting in a total acquisition time of approximately 3.3 h.

X-ray reflectivity (XRR) was carried out using a Philips PW1710 diffractometer equipped with separate θ and 2θ drives, and a line focus Cu-K$_\alpha$ lab-source fitted with a Ni β-filter. A ¼° divergence slit and a 5 mm mask was used on the primary side while a 1 mm antiscatter slit and a ¼° receiving slit was used on the secondary side followed by a secondary Ge 220 monochromator and a gas filled proportional detector.

Neutron reflectivities from $^{10}$B$_4$C/$^{11}$B$_4$C multilayers were predicted using the GenX reflectivity simulation software which uses the Parratt recursion formalism. The structural parameters of as-grown multilayers were deduced by fitting simulated NR and XRR to experimental data using models created within GenX.[24] The neutron scattering length densities were calculated based on the tabulated values for B$_4$C density[25] (ρ=2.52 g/cm$^3$) and neutron scattering lengths for the boron isotopes[26] and carbon.[27]

X-ray diffraction (XRD) was performed using a Panalytical X'Pert Bragg-Brentano θ-θ diffractometer with Cu-K$_\alpha$ X-rays. On the primary side a Bragg-Brentano HD mirror was used with a ½° divergence slit and a ½° antiscatter slit, and on the secondary side a 5 mm antiscatter slit was used together with an X'celerator detector operating in scanning line mode. Diffraction measurements were performed in the range 10°-90° 2θ with a step size of 0.033°/step and a time per step of 50 s, leading to a total acquisition time of approximately 16 min.

An analytical Tecnai G2 UT FEG microscope operating at 200 kV for a point-to-point resolution of 0.19 nm was used for transmission electron microscopy (TEM) studies. A $^{10}$B$_4$C/$^{11}$B$_4$C multilayer with N=10 periods of nominally Λ=128 Å was studied in cross-section using bright field imaging, high resolution TEM, as well as selected area electron diffraction (SAED). The cross-sectional samples were prepared using mechanical grinding and polishing followed by low-energy ion-beam milling using a Gatan precision ion polishing system.

**Results and discussion**

The potential of utilizing $^{10}$B/$^{11}$B modulated boron carbide as multilayer neutron mirrors was confirmed by GenX simulations of NR from 128 nm thick $^{10}$B$_4$C/$^{11}$B$_4$C multilayer mirrors with different bilayer thicknesses Λ. Figure 2 shows the simulated reflectivity of neutrons with wavelength λ=4.85 Å from two multilayer mirrors consisting of 10 bilayers of Λ=128 Å and 20 bilayers of Λ=64 Å, respectively, assuming equal $^{10}$B$_4$C and $^{11}$B$_4$C individual layer thicknesses and interface widths σ=3 Å. As can be seen, both simulations predict very high peak reflectivities in the first multilayer Bragg peaks, despite the limited number of constituting $^{10}$B$_4$C/$^{11}$B$_4$C bilayers. This indicates that $^{10}$B/$^{11}$B isotope modulation within a boron carbide matrix can potentially achieve high reflectivity neutron multilayer mirrors, provided that the interface width is kept in the order of 3 Å. Our simulations also revealed an effect of neutron absorption in the $^{10}$B-containing layers. It will manifest itself primarily as a reduced reflectivity at very grazing incidence angles, i.e., for the first Bragg peaks from multilayers with very large periodicities. This is because at grazing incidence, when the neutron beam path through $^{10}$B-containing layers becomes very long, there will be a trade-off between the number of neutrons that are reflected before they are absorbed. This implies $^{10}$B/$^{11}$B isotope modulated multilayers are most suited for band-pass mirrors and medium Q monochromators and that devices operating at grazing incidence need to be optimized using design rules minimizing the thickness of $^{10}$B-containing layers close to the surface.

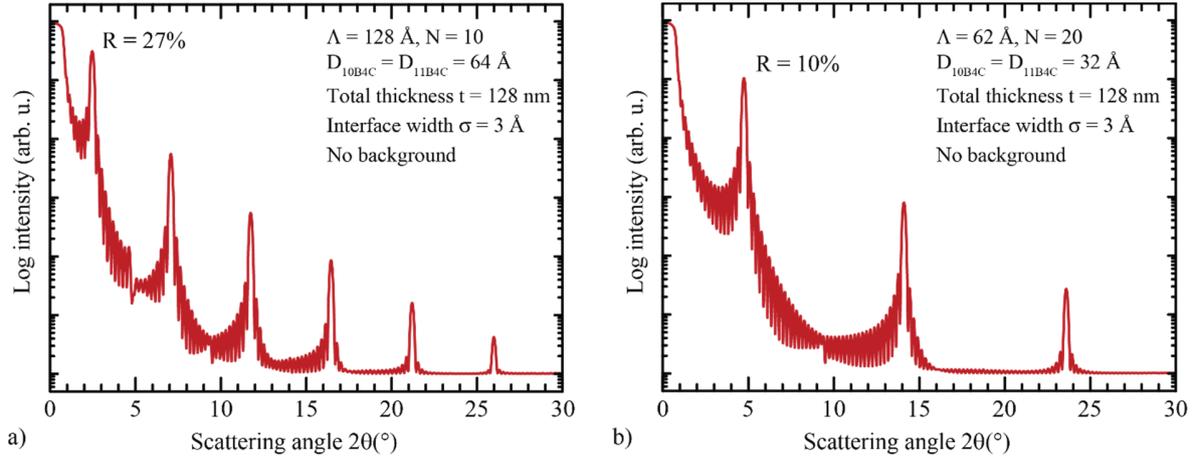

**Figure 2.** a) and b). Simulated neutron reflectivities for two 128 nm-thick $^{10}B_4C/^{11}B_4C$ multilayer mirrors with bilayer periods $\Lambda$=128 Å and $\Lambda$=64 Å, respectively. The interface widths are set to $\sigma$=3 Å and the neutron wavelength is $\lambda$=4.825 Å.

The elemental and isotopic compositions of two as-deposited mirrors with different nominal $^{10}B_4C$ layer-thickness-to-period ratios, denoted as $\Gamma$, of 0.75 and 0.25, respectively, are listed in Table 1. Both multilayers contain N=20 bilayers with the same nominal bilayer periodicity $\Lambda$=64 Å. As can be seen, both multilayers contain about 84 at.% B and 15 at.% C, i.e., $B_xC$ with x=5.7, which is significantly more boron than 80 at.% as expected from the chemical formula $B_4C$ of the polycrystalline sputter targets. However, magnetron sputter deposited thin films from $B_4C$ targets typically achieve a close to x=4 stoichiometry,[6] even when the deposition plasma is highly ionized[28] as in the present case. Since bulk boron carbide is known to exhibit a stable phase with a wide stoichiometry range of 8-20 at.% carbon[8,9] (12.5$\geq$x$\geq$4) we speculate that the high B content observed in our multilayers is attributed to over stoichiometric target materials. It should be noted that a high content of B is beneficial for the application of neutron mirrors owing to an increased SLD contrast at the interfaces with an increasing content of $^{10}B$ and $^{11}B$ which, in turn, increases the neutron reflectivity.

The $^{10}B/(^{10}B+^{11}B)$ isotopic relative compositions exhibit slightly lower values than expected by the designed layer thicknesses (69 at.% vs. 75% for sample ML-075, and 24 at.% vs. 25% for sample ML-025). A likely reason for these discrepancies is that actual $^{10}B_xC$ layers may be thinner than intended, possibly combined with a lower isotope enrichment in the $^{10}B_xC$ target than in the $^{11}B_xC$ target.

**Table 1.** Isotopic and elemental average compositions of two $^{10}B_4C/^{11}B_4C$ multilayer mirrors, "ML-075" and "ML-025", with nominal multilayer periods $\Lambda$=64 Å and $^{10}B_4C$ layer thickness-to-period ratios $\Gamma$ being 0.75 and 0.25, respectively. Nominal values denoted by *.

| | Sample design | | Average composition in multilayer | | | | | Atomic ratios | |
|---|---|---|---|---|---|---|---|---|---|
| Sample | Period* $\Lambda$ | $^{10}B_4C/\Lambda$ thickness ratio* $\Gamma$ | $^{10}B$ (at.%) | $^{11}B$ (at.%) | $^{10}B+^{11}B$ (at.%) | C (at.%) | H + O (at.%) | B/C ratio | $^{10}B/(^{10}B+^{11}B)$ ratio |
| ML-075 | 64 Å | 0.75 | 59.0 | 25.8 | 84.8 | 14.6 | 0.6 | 5.8 | 69% |
| ML-025 | 64 Å | 0.25 | 20.3 | 64.0 | 84.3 | 15.1 | 0.6 | 5.6 | 24% |

In order to investigate the definition of the $^{10}B/^{11}B$ modulation within the boron carbide matrix, neutron reflectivity was performed on the ML-075 sample consisting of 20 repetitions of $^{10}B_xC+^{11}B_xC$ bilayer period with a nominal periodicity of $\Lambda$=64 Å and a nominal $^{10}B_xC$ layer thickness-to-period ratio of $\Gamma$=0.75 (see Table 1). The blue curve in Figure 3 (a) shows the experimental NR data from the mirror where a clear first Bragg peak with an absolute reflectivity of 13% is detected at a scattering vector $Q_Z$=0.206 Å$^{-1}$, which corresponds to a periodicity $\Lambda$=61.5 Å. The red curve shows the best fit by GenX simulations, which was modelled using the values of x=5.7 and $\Gamma$=0.69, as determined by ToF-ERDA. Due to a limited Q-range for this measurement, the fitting did not converge to any unique value of the interface width $\sigma$, but rather to a range of 0-5 Å. This range is consistent with previous work on amorphous multilayers, where the modulated ion-assistance scheme has successfully been applied on amorphous $^{11}B_4C$ containing Ni/Ti based multilayers with interface widths ranging between 2.7 Å and 4.5 Å[17,29,30]. Even smaller interface widths are not unlikely in these $^{10}B_xC/^{11}B_xC$ multilayers thanks to the extremely strong covalent bonding[8] which minimize the adatom mobility during the initial 0.3 nm non-ion-assisted growth, in combination with $B_xC$'s insensitivity to uncontrolled kinetic intermixing effects. Thus, given the high local contrast in terms of SLD, these interference mirrors show a strong potential for neutron optical components.

Figure 3 (b) shows an X-ray reflectivity measurement (blue curve) performed to reveal if any chemical modulation exist in an N=10 multilayer with nominal Λ=128 Å and Γ=0.5. The red curve shows the best fit obtained by GenX simulations using the nominal multilayer parameters with an interface width σ=3 Å plus a 19.6 Å low-density oxide top layer. The vertical dashed lines indicate the positions for multilayer Bragg peaks in case of any 128 Å chemical modulation existing in the sample. As can be seen, this is not the case.

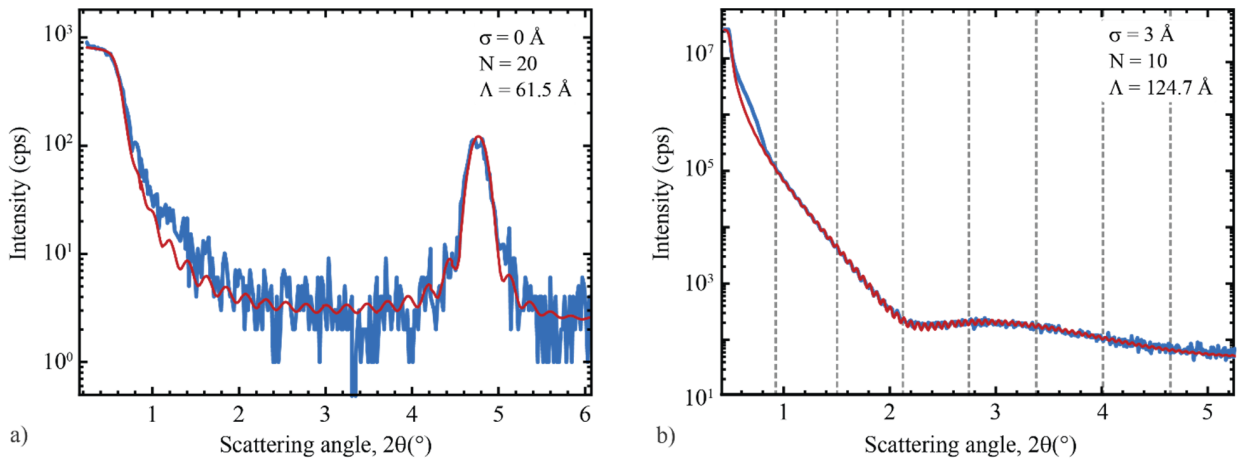

**Figure 3.** (a) Neutron reflectometry measurement of a $^{10}B_{5.7}C/^{11}B_{5.7}C$ multilayer with N=20 bilayers and Λ=61.5 Å. A clear multilayer peak occur at approximately 4.7° for λ=4.825 Å. The best fit GenX simulations revealed $^{10}B_{5.7}C$ and $^{11}B_{5.7}C$ layer thicknesses equal to 42.5 Å and 19 Å, respectively. (b) X-ray reflectometry measurement of a $^{10}B_4C/^{11}B_4C$ multilayer consisting of N=10 bilayers of nominal modulation period Λ=128 Å. The dashed lines indicate the positions for where multilayer Bragg peaks would appear if a 128 Å chemical modulation had been present in the sample. Blue and red curves in (a) and (b) represent experimental data and simulation, respectively.

The microstructure and possible chemical modulation were further investigated using transmission electron microscopy. Figure 4 (a) is a bright field TEM image showing an overview of the entire multilayer stack of N=10 repetitions of bilayer period Λ=128 Å. The inset shows a SAED pattern obtained from the entire multilayer stack. It was not possible to discern any sign of the $^{10}B_{5.7}C/^{11}B_{5.7}C$ modulation confirming that the multilayer is chemically homogenous. Figure 4 (b) shows a high resolution image from the first two periods of the multilayer which reveals an amorphous structure with no signs of the $^{10}B/^{11}B$ interfaces. These structural findings were corroborated by the X-ray diffraction analysis (not shown) thus confirming that the isotope modulated multilayers are amorphous and chemically homogeneous boron carbide also over large areas which can not be measured by TEM.

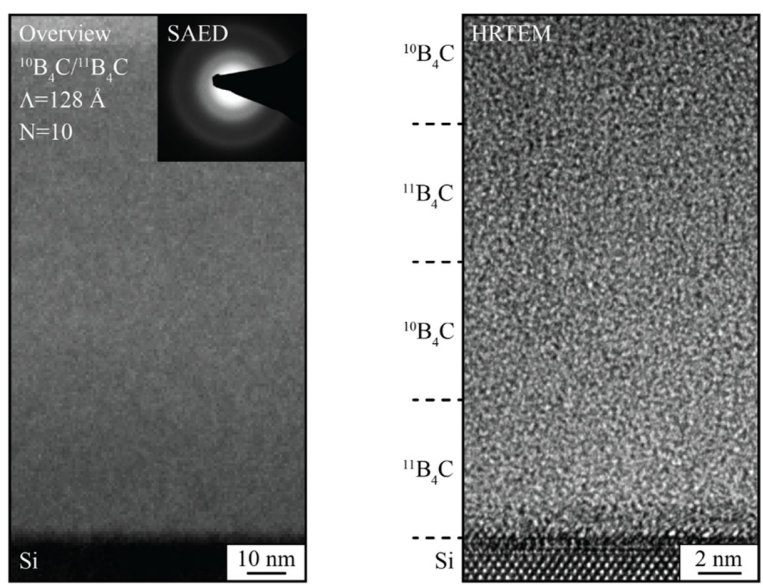

**Figure 4.** TEM of a $^{10}B_4C/^{11}B_4C$ multilayer with N=10 repetitions of bilayer period Λ=128 Å. (a) Shows an overview of the entire 10 periods in the film. The inset shows a SAED pattern obtained from the entire multilayer stack. (b) Shows a high resolution image from the first two periods of the multilayer.

## Conclusions

Motivated by our idea of improving the interface widths in multilayer neutron mirrors by eliminating any chemical driving force for deteriorating interface abruptness, we have for the first time investigated the feasibility of utilizing the high SLD contrast provided by $^{10}B/^{11}B$ isotope-modulation only, for implementation as multilayer neutron mirrors. Model simulations of isotope modulated $^{10}B_4C/\ ^{11}B_4C$ multilayers clearly indicate that a high neutron reflectance will be achievable even for a very limited number of bilayers with periodicities smaller than Λ<128 Å and an interface width of σ=3 Å. We have synthesized a series of multilayers by alternating magnetron sputter deposition from $^{10}B_4C$ and $^{11}B_4C$ targets, utilizing a low-energy high-flux modulated ion assistance in order to achieve as abrupt interfaces as possible. Elemental and isotopic compositional analyses show that the multilayers exhibit a boron-to-carbon number density ratio x=5.7 which, for an unknown reason, is considerably higher than the nominally stoichiometric $B_4C$ sputtering targets. Neutron reflectivity measurements of a $^{10}B_{5.7}C/^{11}B_{5.7}C$ multilayer with N=20 and Λ=61.5 Å exhibits a first Bragg peak of 13% reflectivity at a scattering angle of 4.7° which confirms the high neutron reflectivity predicted by the initial model simulations. Fitting of neutron reflectivity simulations clearly show that a $^{10}B/^{11}B$ interface width s is in the range $0<\sigma\leq 5$ Å, which is better that today's state-of-the-art Ni/Ti neutronmirrors. The higher than expected boron content may play a role for achieving high neutron reflectivity because the boron isotopes are less diluted with carbon for higher x which, in turn, increases the SLD contrast in the multilayers. The boron carbide matrix, hosting the $^{10}B$ and $^{11}B$ isotopic layers, is shown to be amorphous without any detectable chemical modulation, as seen by TEM, SAED and XRD.

Our findings open new possibilities for bandpass multilayer mirrors and medium q multilayer monochromators exhibiting high reflectivities from $^{10}B_{5.7}C/^{11}B_{5.7}C$ multilayers consisting of a relatively low number of layers.

## Acknowledgments

This research was funded by the Swedish Foundation for Strategic Research (SSF) within the Swedish national graduate school in neutron scattering (SwedNess), and the Swedish Research Council (VR). The Swedish Research Council VR Grant numbers 2019-00191 and 2021-00357 (for accelerator-based ion-technological center in tandem accelerator laboratory in Uppsala University) and the Swedish Government Strategic Research Area in Materials Science on Advanced Functional Materials (AFM) at Linköping University (Faculty Grant SFO Mat LiU No. 2009 00971) are also acknowledged. The authors would also like to acknowledge Jochen Stahn for his help with the neutron reflectivity measurements at the MORPHEUS beamline of the SINQ facility at the Paul Scherrer Institute.